\documentclass[aps,
prd,
10pt,
preprint,
tightenlines,
superscriptaddress,
nofootinbib,
showpacs
]{revtex4-1}
\usepackage{amssymb,latexsym}
\usepackage{amsmath,amsbsy,bbm}
\usepackage{epsfig,bm}
\usepackage{graphicx,comment}
\begin{document}
\def\a{{\alpha}}
\def\b{{\beta}}
\def\d{{\delta}}
\def\D{{\Delta}}
\def\X{{\Xi}}
\def\e{{\varepsilon}}
\def\g{{\gamma}}
\def\G{{\Gamma}}
\def\k{{\kappa}}
\def\l{{\lambda}}
\def\L{{\Lambda}}
\def\m{{\mu}}
\def\n{{\nu}}
\def\o{{\omega}}
\def\O{{\Omega}}
\def\S{{\Sigma}}
\def\s{{\sigma}}
\def\th{{\theta}}

\def\ol#1{{\overline{#1}}}

\def\cF{{\mathcal F}}
\def\cS{{\mathcal S}}
\def\cC{{\mathcal C}}
\def\cB{{\mathcal B}}
\def\cT{{\mathcal T}}
\def\cQ{{\mathcal Q}}
\def\cL{{\mathcal L}}
\def\cO{{\mathcal O}}
\def\cA{{\mathcal A}}
\def\cQ{{\mathcal Q}}
\def\cR{{\mathcal R}}
\def\cH{{\mathcal H}}
\def\cW{{\mathcal W}}
\def\cM{{\mathcal M}}
\def\cD{{\mathcal D}}
\def\cN{{\mathcal N}}
\def\cP{{\mathcal P}}
\def\cK{{\mathcal K}}

\def\eqref#1{{(\ref{#1})}}

\preprint{RBRC-963}

\title{A Method to Extract Charged Hadron Properties from Lattice QCD in Magnetic Fields}

\author{B.~C.~Tiburzi}
\email[]{btiburzi@ccny.cuny.edu}

\affiliation{Department of Physics, The City College of New York, New York, NY 10031, USA}
\affiliation{Graduate School and University Center, The City University of New York, New York, NY 10016, USA}
\affiliation{RIKEN BNL Research Center, Brookhaven National Laboratory, Upton, NY 11973, USA}

\author{S.~O.~Vayl}
\email[]{stevenvayl43@gmail.com}
\affiliation{Department of Physics, The City College of New York, New York, NY 10031, USA}
\affiliation{Graduate School and University Center, The City University of New York, New York, NY 10016, USA}

\date{\today}


\begin{abstract}
By analyzing the external field dependence of
 correlation functions,
the magnetic properties of hadrons can be determined using lattice QCD in magnetic fields.
To compute the magnetic moments and polarizabilities of charged hadrons, 
for example, 
one requires sufficiently weak magnetic fields. 
Such field strengths, 
however, 
lead to closely spaced Landau levels that are not straightforwardly resolved using standard lattice spectroscopy.
Focusing on charged spinless hadrons, 
we introduce a simple projection technique that can be used to isolate the lowest Landau level. 
As the technique requires the explicit coordinate-space wave-function, 
we investigate the extent to which the continuum, infinite volume wave-function can be employed.
We find that, 
in practice,  
the effects of discretization can be handled using a perturbative expansion about the continuum. 
Finite volume corrections are taken into account by using the discrete magnetic translational invariance of the torus. 
We show that quantized magnetic fields can lead to pernicious volume effects which depend on the magnetic flux quantum, 
rather than on the lattice volume.  
\end{abstract}

\maketitle

\section{Introduction}%

Quarks are unique in the Standard Model because they are the only particles charged under all three gauge groups. 
While electromagnetic interactions of quarks are dwarfed by their strong interactions, 
electromagnetic observables and tiny electrodynamic effects ultimately 
give one an intuitive picture of the underlying QCD dynamics present in the vacuum and in hadrons.
Studying QCD in the presence of classical electromagnetic fields, 
moreover, 
allows one to address how QCD responds to external conditions.
In recent years, 
considerable steps have been taken to investigate QCD in external electromagnetic fields using lattice gauge theory techniques;
for an overview, 
see
\cite{Tiburzi:2011vk,D'Elia:2012tr}.

The external field method in lattice QCD appeared quite early on. 
Background magnetic fields were used in the first calculations of the nucleon magnetic moments%
~\cite{Martinelli:1982cb,Bernard:1982yu}.
Soon thereafter, 
it was realized that effects at second order in background electric fields could be used to extract the electric polarizabilities of neutral hadrons%
~\cite{Fiebig:1988en}. 
In the context of weak external magnetic fields, 
the background field method was revisited several years ago to compute magnetic moments and polarizabilities of hadrons%
~\cite{Lee:2005ds,Lee:2005dq}.
These studies were completely quenched, and non-uniform magnetic fields were employed with effects resulting from field gradients mitigated by the imposition of spatial Dirichlet boundary conditions.
Such boundary conditions produce non-perturbative effects that are unfortunately difficult to quantify.

Uniform magnetic fields can be achieved on the lattice by imposing a quantization condition
on magnetic field strength%
~\cite{'tHooft:1979uj,Smit:1986fn,Damgaard:1988hh}.
In QCD, 
this condition takes the form
\begin{equation} \label{eq:quantization}
|q_d| B 
= 
\frac{
2 \pi n_\Phi
}
{ 
L^2   
}
,\end{equation}
where 
$L$ 
is the spatial size of the lattice, 
which we assume to be the same in each spatial direction, 
and 
$|q_d| = \frac{1}{3} e$ 
is the magnitude of the electric charge of the down quark. 
The integer
$n_\Phi$
is the magnetic flux quantum of the torus. 
A study of magnetic moments of spin-$3/2$ resonances employing uniform magnetic fields appeared a few years ago%
~\cite{Aubin:2008qp}.
While the quantization condition has generally restricted one to prohibitively large magnetic fields, 
lattice volumes are ever increasing, and this trend will consequently lead to lattices that are able to support several quantized values of perturbatively small magnetic fields.

Due to the existence of Landau levels, 
the correlation functions of charged hadrons depend strongly on the magnetic field.  
For a hadron of mass 
$M$, 
the characteristics of the Landau levels depend on the dimensionless ratio
$| e B | / M^2$.  
When the magnetic field is large, 
$| e B | / M^2 \gg 1$, 
the Landau levels are widely separated in energy. 
In this regime, 
one can expect that standard lattice spectroscopy will filter out the lowest Landau level in the long Euclidean time limit. 
On the other hand, 
for small values of the magnetic field, 
$| e B | / M^2 \ll 1$, 
the energies of Landau levels are very closely spaced and standard spectroscopic techniques will be of limited use. 
To deduce the magnetic properties of charged hadrons,%
\footnote{
As these properties are dynamically determined, 
one must additionally worry about energy scales set by the virtual degrees of freedom within hadrons. 
Because the charged pion is the lightest electrically charged state in QCD, 
the ratio
$| e B | / m_\pi^2$ 
is important for all hadrons. 
} 
one is ultimately interested in the regime 
$| e B | / M^2 \ll 1$, 
however, 
in the near term one may have to settle for more intermediate values, 
for which
$| e B | / M^2 \lesssim 1$. 
Nonetheless, 
as the available magnetic fields become smaller, 
the need to explicitly treat Landau levels becomes greater.
Fortunately this can be achieved easily using the proper-time formalism, 
which was pioneered in quantum field theory by Schwinger%
~\cite{Schwinger:1951nm}.
In this work, 
we develop the technique for lattice correlation functions by focusing on charged scalar particles.

In an external magnetic field with 
$| e B | / M^2 \lesssim 1$, 
the tower of Landau levels complicates standard lattice spectroscopy. 
As we show, 
one need not rely upon the long-time limit of the correlation function to filter out the lowest Landau level. 
One can project out this state directly, 
and thereby considerably reduce the challenges inherent to studying charged
hadrons in external magnetic fields. 
We begin in 
Sec.~\ref{s:method}
by detailing the method using a continuum action defined in infinite volume. 
For simplicity, 
we focus on charged scalar particles. 
To address the effects of discretization, 
we consider a charged particle effective action on an infinite lattice in 
Sec.~\ref{s:discrete}. 
We find that the effects of discretization can be treated in practice using perturbation theory with a continuum 
wave-function. 
Next in Sec.~\ref{s:FV}, 
we investigate finite volume corrections.
Using the discrete magnetic translational invariance of the torus, 
we account for finite volume corrections to the lowest Landau level. 
While volume corrections are generally suppressed by exponentially small terms, 
we expose certain corrections that are not exponentially small, 
and discuss how the infinite volume limit can be recovered. 
This subtlety owes to the magnetic field quantization required on the torus. 
A simple numerical study is undertaken in 
Sec.~\ref{s:test}, 
where we compute various two-point functions for a scalar particle coupled to a magnetic field on a finite, 
four-dimensional lattice. 
We compare the numerical computations to the analytic results we derive, 
and find excellent agreement. 
Furthermore the numerical results show that the projection technique is necessary for magnetic fields that are 
perturbatively small compared to the hadron's mass scale. 
Finally we conclude in 
Sec.~\ref{s:summy}
with a brief summary of our findings, 
and directions for future work.

\section{Continuum Method} \label{s:method}  %

The object for studying the spectrum of QCD is the two-point correlation function. 
Standard lattice spectroscopy relies on the large Euclidean time limit of the two-point correlator.
With a mixed momentum-time representation, 
the two-point correlator is%
\footnote{
Throughout we omit explicitly writing the end-points of integration for all integrals spanning the entire real line. 
Thus we write
$\int dx$
for the integral
$\int_{-\infty}^{+\infty} dx$. 
Similarly for sums over all integers,
we do not explicitly write out the limits. 
Thus we use the notation
$\sum_n$ 
for the sum
$\sum_{n = - \infty}^{+\infty}$. 
}
\begin{equation} \label{eq:ptau}
G(\vec{p}, \tau) 
= 
\int d\vec{x}  \, e^{ i \vec{p} \cdot \vec{x}}
\langle 0 | \Phi(\vec{x}, \tau) \Phi^\dagger(\vec{0}, 0) | 0 \rangle
,\end{equation}
where for simplicity we restrict our attention to a charged scalar hadron 
$\phi$
which has an interpolating field
$\Phi$. 
The Fourier transform introduces the coordinate-space wavefunction, 
$\psi_{\vec{p}}^* (\vec{x}) = e^{ i \vec{p} \cdot \vec{x}}$,
 of the momentum eigenstate 
$\vec{p}$. 
The spectral decomposition of the correlator
shows that the ground state saturates the correlation function in the long Euclidean time limit,
with excited-state contamination suppressed by the factor
$\sim \exp( - \D E \, \tau)$. 
Here 
$\D E$
is the splitting between the ground and first excited states, 
and we have assumed that the overlap factors for both states are approximately equal. 
For Euclidean times on the order of 
$\tau \sim 1 \, \texttt{fm}$, 
a natural energy-level splitting, 
$\D E \sim \Lambda_{\text{QCD}}$, 
ensures reasonably good suppression of the excited states.

The addition of a magnetic field, 
$\vec{B}$,
considerably complicates the spectral decomposition of the two-point function. 
Energy levels of multiparticle states, 
for example, 
become cumbersome to enumerate. 
We will assume the zero-field ground state is the single-hadron state 
$\phi$. 
This ground-state hadron, 
which is a state possessing the quantum numbers of 
$\Phi$,
now consists of an infinite tower of Landau levels. 
These energy levels are not widely separated in weak magnetic fields.
For an ideally weak external field, 
the energy splitting between neighboring Landau levels is set by
$\D E = | e B | / M$, 
where 
$M$ 
is the hadron's mass. 
In a given magnetic field, 
the proton's Landau levels, 
for example, 
will be more closely spaced than the charged pion's leading to comparatively more challenging lattice spectroscopy. 
Even for the best case,
the pion itself,  
the energy level splitting in perturbatively weak fields requires 
$\tau$
considerably greater than 
$1 \, \texttt{fm}$
to resolve. 
Consequently it is beneficial to handle the Landau levels explicitly.

The single-particle effective action for the composite field
$\phi$
in a uniform electromagnetic field has the form
\begin{eqnarray} \label{eq:effact}
S
&=&
\int d^4 x
\left[
D_\mu \phi^\dagger D_\mu \phi 
+ 
M^2 \phi^\dagger \phi 
+
c_0
F_{\mu \nu} F_{\mu \nu} 
\phi^\dagger \phi
+ 
c_2
F_{\rho \{ \mu} F_{\nu \} \rho}
D_\mu \phi^\dagger D_\nu \phi
\right]
,\end{eqnarray}
which follows on account of gauge invariance, parity invariance, and Euclidean invariance. 
The gauge covariant derivative is defined in the usual way, 
$D_\mu \phi = \partial_\mu \phi + i e A_\mu \phi$,
and the electromagnetic field-strength tensor is 
$F_{\mu \nu} = \partial_\mu A_\nu - \partial_\nu A_\mu$.
The curly braces denote the symmetric traceless part of a tensor, 
namely
$\cO_{\{ \mu \nu \}} 
= 
\frac{1}{2} 
\left( 
\cO_{\mu \nu} 
+ 
\cO_{\nu \mu} 
- 
\frac{1}{2} 
\d_{\mu \nu} \cO_{\rho \rho} 
\right)$.
The combination 
$T_{\mu \nu}= F_{\rho \{ \mu} F_{\nu \} \rho}$
is the electromagnetic stress-energy tensor. 
In the effective action, 
we have written down all terms containing up to two powers of the electromagnetic field-strength tensor. 
There are of course higher-order terms allowed by symmetries, 
but we assume the external field to be sufficiently weak. 
The low-energy constants, 
$c_0$ 
and 
$c_2$,
describe the helicity, 
$\lambda = 0$ and $\lambda = 2$, 
couplings of two photons to the 
$\phi$. 
From a simple matching calculation, 
these low-energy constants can be identified as
\begin{eqnarray}
c_0
&=&
\pi M ( \alpha_E - \beta_M),
\quad
c_2 
=
\frac{4 \pi}{M} ( \alpha_E + \beta_M)
,\end{eqnarray}
where 
$\alpha_E$
is the electric polarizability of the 
$\phi$, 
and 
$\beta_M$
its magnetic polarizability.

Now we specialize to the case of a magnetic field specified by the vector potential%
\footnote{ 
The charged particle correlation function defined in 
Eq.~\eqref{eq:dumb}
is gauge variant,
as are the generalizations defined in Eqs.~\eqref{eq:Btau}, \eqref{eq:Btaulatt}, and \eqref{eq:coord}. 
To remove the gauge dependence, 
one could alternatively include a Wilson line between the field operators to form gauge invariant two-point correlation functions. 
This approach necessitates choosing a path linking the space-time locations of the field operators. 
The Wilson line is not independent of the path, 
however,
because magnetic flux threads closed loops transverse to the field direction. 
This dependence was explored previously in a lattice calculation by choosing different definitions of the charged particle two-point function%
~\cite{Rubinstein:1995hc}.
In the present method, 
the effective action is used to compute the correlation functions, 
and we have chosen a convenient gauge to accomplish this. 
In a different gauge, 
the behavior of the correlation function will be different, 
and consequently the coordinate wave-function used to project out the lowest Landau level will differ. 
The virtue of the effective action approach is that in each gauge there is a prediction for the behavior of the correlation function. 
We can generalize the effective action approach, 
moreover, 
for gauge invariant charged particle correlation functions to study the path dependence introduced by the Wilson line. 
To keep our presentation simple, 
we introduce the method using gauge variant correlators restricted to the gauge specified by Eq.~\eqref{eq:A}. 
}
\begin{equation} \label{eq:A}
A_\mu = ( - B x_2, 0, 0, 0)
.\end{equation}
After a field redefinition, 
the effective action can be written in the form
\begin{equation} \label{eq:effective}
S 
= 
\int d^4 x 
\left[
D_\mu \phi^\dagger D_\mu \phi + \cM^2 \phi^\dagger \phi
\right]
,\end{equation}
%
\newpage
%
\noindent
where 
$\cM$ 
is given by
\begin{equation} \label{eq:M}
\cM(B) = M - \frac{1}{2} 4 \pi \beta_M B^2 + \cO(B^4)
,\end{equation}
and the higher-order terms in the magnetic field that were neglected above are now effectively subsumed into the 
$B$-dependence of 
$\cM$. 
The goal of a background field lattice computation is to determine the magnetic polarizability 
$\beta_M$
from the magnetic field dependence of 
$\cM$. 

In an external magnetic field,
one can consider the two-point correlator, 
\begin{equation}  \label{eq:dumb}
G_B (\tau)
=
\int d \vec{x} \,
\langle 0 | \Phi(\vec{x},\tau) \Phi^\dagger (\vec{0}, 0) | 0 \rangle_B
,\end{equation}
where the subscripts denote the external field dependence. 
Because the 
$x_2$-component of momentum is not a good quantum number in the external magnetic field, 
the Fourier transform implicit in Eq.~\eqref{eq:dumb}
receives contributions from an infinte tower of Landau levels. 
Using the effective action in Eq.~\eqref{eq:effective}
and following%
~\cite{Tiburzi:2008ma},
the 
$\phi$
contribution to the correlator can be written in the form%
\footnote{
The overlap factor, 
which is generically written as 
$Z_\phi$,
also contains kinematic factors.  
We will write the same overlap factor 
$Z_\phi$
throughout; 
although, 
it may have different numerical values in different correlators. 
Because the overlap of the interpolating field 
$\Phi$
with the single particle state 
$\phi$
is \emph{a priori} unknown, 
we do not keep track of such kinematic factors. 
}
\begin{equation} \label{eq:mess}
G_B(\tau)
=
Z_\phi
\int_0^\infty ds
\frac{e^{- \frac{1}{2 s} \left( \tau^2 + s^2 \cM^2 \right)}}{\sqrt{s \cosh ( e B s)}}
,\end{equation}
where the non-standard 
$\tau$-dependence 
arises from summing contributions from the entire tower of Landau levels. 
In the extreme long-time 
and non-relativistic limits, 
only the lowest Landau level contributes,
$G_B(\tau) \sim \exp \left[ - (\cM + \frac{|e B|}{2 \cM} )\tau \right]$.
As lattice calculations are far from these ideal limits, 
one must confront contributions from higher Landau levels. 
The explicit form of the correlation function in Eq.~\eqref{eq:mess} could be used, 
in principle, 
to fit the lattice correlator,%
\footnote{
The utility of external field correlators in fitting lattice data was originally suggested for background electric field computations~\cite{Detmold:2006vu}. 
Analogous to the Landau level problem, 
charged particle two-point functions in external electric fields have non-standard 
$\tau$-dependence, 
and the predicted functions have been sucessfully used to extract properties
of pseudo-scalar mesons and baryons~\cite{Detmold:2009dx,Detmold:2010ts}. 
} 
however, 
there are hidden assumptions. 
In deriving this expression for the correlator, 
we integrated over all space. 
This leads to contributions from Landau levels with an average size extending a considerable distance from the origin. 
Such contributions are modified due to boundary conditions.
It is thus highly desirable to filter out only the lowest Landau level 
which has the smallest average size.

To remedy the situation with narrowly separated energy levels, 
we return to the correlation function in 
Eq.~\eqref{eq:dumb}. 
By using this form, 
one is implicitly assuming the single-particle wave-function,
$\psi_{p_2}^* (x_2) = e^{ i p_2 x_2}$,
of a free particle with 
$p_2 =0$,
\footnote{
If one imposes spatial Dirichlet boundary conditions, 
as is done in some external field lattice calculations, 
see~\cite{Freeman:2012cy}, 
for example, 
there is an analogous issue even for neutral particle correlation functions. 
The sum over spatial lattice sites produces a tower of standing waves, 
and practitioners rely on the long-time limit to obtain the lowest Dirichlet mode from the tower. 
This tower could be directly avoided, 
however, 
by projecting out the lowest mode using its coordinate wavefunction, 
$\psi (x) = \sin \left( \frac{\pi x}{L} \right)$.
While this suggestion would circumvent the momentum problem, 
it unfortunately does nothing to resolve the myriad difficulties with Dirichlet boundary conditions, 
e.g.~the issue of the 
$U(1)_A$ 
anomaly~\cite{Atiyah:1975jf,Ninomiya:1984ge}, 
the possibility of chiral symmetry restoration~\cite{Flachi:2012pf,Flachi:2013bc,Tiburzi:2013vza}, 
\emph{etc}. 
} 
\emph{cf}.~Eq.~\eqref{eq:ptau}.
A more desirable correlation function is
%
\newpage
%
\begin{equation} \label{eq:Btau}
\mathcal{G}_B(\tau)
=
\int d \vec{x} \,
\psi^{* (0)}_0(x_2) \,
\langle 0 | \Phi(\vec{x},\tau) \Phi^\dagger (\vec{0}, 0) | 0 \rangle_B
,\end{equation}
where 
$\psi^{(0)}_0(x)$ 
is the ground-state harmonic oscillator wave-function
\begin{equation} \label{eq:Gauss}
\psi^{(0)}_0(x) 
= e^{ - \frac{1}{2} | e B | x^2}
.\end{equation} 
The absolute normalization of this wave-function is irrelevant for our purposes, 
and the notation for the wave-function is chosen to be consistent with Sec.~\ref{s:discrete} below.  
Computing the 
$\phi$
contribution to the correlator in 
Eq.~\eqref{eq:Btau}
produces a simple result. 
Using Schwinger's proper-time trick, 
we find an exponential fall-off of the correlator
\begin{equation} \label{eq:LLL}
\mathcal{G}_B(\tau)
=
Z_\phi \,
e^{-  E_0 \tau }
,\end{equation}
with the energy given by
\begin{equation} \label{eq:E0}
E_0 
= 
\sqrt{\cM^2 + |e B|}
.\end{equation}
The new correlation function 
$\mathcal{G}_B(\tau)$
allows one clean access to 
$\cM$, 
and consequently to the magnetic polarizability 
$\beta_M$.

An obvious drawback of the projection method is that the coordinate-space wave-function for the lowest Landau level is required. 
On a finite space-time lattice, 
this wave-function does not have the simple Gau{\ss}ian form well-known from elementary quantum mechanics.  
Modifications to the wave-function will arise from the discretization of space, 
as well as the finite spatial volume. 
We consider each of these effects below.

\section{Discretization Effects} \label{s:discrete} %

We investigate first the effects of discretization on the lowest Landau level. 
To this end, 
we temporarily ignore the boundary conditions.
Addressing finite-size effects is postponed until the following section. 
In the absence of boundary conditions, 
a natural question remains. 
Can one reasonably approximate the lowest Landau level using a continuum wave-function? 
To investigate the answer to this question, 
we consider the theory on an infinite spatial lattice with a uniform lattice spacing 
$a$.
The lattice sites are labeled by a vector of integers,
$\vec{n} = (n_1, n_2, n_3)$, 
which corresponds to the coordinate vector
$\vec{x} = a \vec{n}$. 
For simplicity, 
we keep the time direction continuous.

The projection method obviously generalizes to the lattice, 
where one now requires the lowest lattice Landau level. 
It is described by a wave-function, 
$\psi_{0, n_2}$, 
where we use the subscript
$n_2$ 
to index the lattice-site dependence.
Generally we use subscripts for discrete indices, 
for example, 
the lattice version of the interpolating field 
$\Phi(x)$
is now written
$\Phi_{\vec{n}}(x_4)$. 
With the wave-function,
$\psi_{0, n_2}$,  
we then form the discrete analogue of Eq.~\eqref{eq:Btau}, 
\begin{equation} \label{eq:Btaulatt}
\mathcal{G}_B (\tau)
= 
\sum_{\vec{n}}
\psi_{0, n_2}^*
\langle 0 | \Phi^{\phantom{\dagger}}_{\vec{n}}(\tau) \Phi {}_{\vec{0}}^\dagger (0) | 0 \rangle_B
.\end{equation}
Consequently
the sum over 
$n_2$ 
in the lattice correlator, 
Eq.~\eqref{eq:Btaulatt},
will project out only the lowest energy state in the tower of lattice Landau levels.

To determine the wave-function for the lowest lattice Landau level,
we must know the lattice form of the single-particle effective action,
which is the discretized version of Eq.~\eqref{eq:effective}.
Using this action, 
one solves the discrete eigenvalue equation for the lowest energy eigenstate, 
and 
$\psi_{0, n_2}$
is then the corresponding eigenvector. 
We will assume a highly plausible form for the discrete action of the charged scalar. 
When one couples gauge fields to matter fields on a lattice, 
gauge invariance requires the use of 
link variables, 
$U_{\mu,x} = \exp ( i e a A_{\mu,x} )$.
A lattice form of the effective action consistent with cubic symmetry, 
gauge invariance, 
and Hermiticity appears as
\begin{eqnarray}
S
&=&
a^3
\sum_{\vec{n}, \vec{n}'}
\int dx_4
\, \phi^\dagger_{\vec{n}} (x_4)
\cD_{\vec{n},\vec{n}'}
\,
\phi_{\vec{n}'} (x_4)
,\end{eqnarray}
with
\begin{eqnarray}
\cD_{\vec{n},\vec{n}'}
&=&
\d_{\vec{n},\vec{n}'}
\left[
-
\frac{\partial^2}{\partial x_4^2} 
+
\cM^2 
\right]
-
\frac{1}{a^2}
\sum_{j=1}^3
\left[
\d_{\vec{n} + \hat{j},\vec{n}'}
\, U_{j,\vec{n}}
+ 
\d_{\vec{n}, \vec{n}' + \hat{j}}
\, U_{j,\vec{n}'}^\dagger
- 
2 
\d_{\vec{n},\vec{n}'}
\right]
.
\label{eq:blah}
\end{eqnarray}
In writing Eq.~\eqref{eq:blah}, 
we have made use of the vanishing of the temporal component of the gauge field, 
and the time-independence of the gauge field. 
In the absence of the external field, 
Eq.~\eqref{eq:blah}
leads to a dispersion relation of the form
\begin{equation} \label{eq:disp}
E^2
=
M^2
+
\frac{4}{a^2}
\sum_{j=1}^{3}
\sin^2 \left( a p_j / 2 \right)
.\end{equation}
One can use the measured spatial momentum dependence of scalar particle energies to establish the precise form of the lattice dispersion relation in a given lattice calculation. 
For our present purposes, 
we assume the dispersion relation to be given by Eq.~\eqref{eq:disp}, 
and the discrete action in Eq.~\eqref{eq:blah} is merely the corresponding gauged action.

Returning to the assumed form for the discretized single-particle action in Eq.~\eqref{eq:blah}, 
we see that the gauge links for the external field,
$U_{j, \vec{n}}$,
introduce explicit lattice-site dependence. 
As a result, 
the 
$x_2$-component of lattice momentum is not a good quantum number. 
This mirrors the situation in the continuum. 
The remaining spatial lattice momenta,
$p_1$ 
and 
$p_3$,
are good quantum numbers, 
as is the time-component of momentum, 
$p_4$.
It is thus useful to take the Fourier transform;
namely, 
in terms of
$\tilde{p}_\mu = (p_1, 0, p_3, p_4)$,
we have
\begin{equation}
\tilde{\phi}_{n_2} (\tilde{p}) 
= 
\sum_{n_1, n_3} 
\int dx_4 \,
e^{ i \tilde{p}_\mu  x_\mu}
\phi_{\vec{n}} (x_4)
.\end{equation}
In terms of the Fourier transformed field, 
the action now appears as
\begin{equation}
S
=
\sum_{n_2,n'_2}
\int_{-\frac{\pi}{a}}^{+\frac{\pi}{a}} \frac{d p_1 d p_3}{(2\pi)^2}
\int \frac{d p_4}{2 \pi}
\tilde{\phi}^\dagger_{n_2} (\tilde{p})
\cD_{n_2,n'_2} (\tilde{p})
\tilde{\phi}_{n'_2} (\tilde{p})
.\end{equation}
Because we can consider correlation functions with two vanishing components of spatial lattice momentum,
$p_1 = p_3 = 0$, 
we restrict our attention to this sector, 
in which
\begin{eqnarray}
\cD_{n_2, n'_2} (p_4)
&=&
T_{n_2, n'_2}
+
\d_{n_2, n'_2}
\left[ 
p_4^2 
+ 
\cM^2  
+
V(n_2) 
\right], \quad
\end{eqnarray}
where the discretized kinetic term is
\begin{equation}
T_{n_2,n'_2}
= 
-
\frac{1}{a^2} 
\left[ 
\delta_{n_2, n'_2 + 1}
+ 
\delta_{n_2 + 1, n'_2}
- 
2
\delta_{n_2, n'_2}
\right]
,\end{equation}
and the potential term is
\begin{equation}
V(n_2)
=
\frac{4}{a^2} \sin^2 \left( e a^2 B n_2  / 2 \right) 
.\end{equation}
The eigenfunctions that diagonalize the operator
$T + V$
are solutions to the lattice Landau level problem. 
The lowest level, 
for example, 
satisfies the coordinate-space eigenvalue equation,
$(T + V) \psi_{0, n_2} = \lambda_0 \, \psi_{0, n_2}$. 
As a consequence, 
the 
$\phi$ 
contribution to the lattice two-point correlation function in Eq.~\eqref{eq:Btaulatt}
has the simple exponential form
\begin{equation}
\mathcal{G}_B(\tau)
=
Z_\phi
e^{ - \mathcal{E}_0 \tau}
,\end{equation}
with the energy eigenvalue given by
\begin{equation} \label{eq:energy}
\mathcal{E}_0 
= 
\sqrt{\cM^2 + \lambda_0}
.\end{equation}

Let us investigate the form of the eigenvalue equation for the lattice Landau level problem more closely. 
In the auxiliary proper-time quantum mechanics, 
the operator 
$T + V$
appears as 
(twice) 
the Hamiltonian, 
and can be written in the abstract form
\begin{equation} \label{eq:lLl}
T + V
= 
\frac{4}{a^2} 
\Big[
\sin^2 \left( a \hat{p}_2 / 2 \right)
+
\sin^2 \left( e a B \hat{x}_2 / 2 \right)
\Big]
.\end{equation}
In the continuum limit, 
we have 
(twice) 
the Hamiltonian of the simple harmonic oscillator
\begin{equation}
T_0 + V_0
= 
\hat{p}_2^2 + e^2 B^2 \hat{x}_2^2
,\end{equation}
which respects the canonical transformation
$(\hat{x}_2, \hat{p}_2) 
\to 
\left( 
\frac{\hat{p}_2}{| e B |}, 
| e B | \hat{x}_2
\right)$.
The lattice Landau level problem, 
Eq.~\eqref{eq:lLl}, 
also respects this canonical transformation.

The dimensionless parameter governing the expansion of the Hamiltonian in 
Eq.~\eqref{eq:lLl}
about the continuum limit is the magnetic field in lattice units, 
namely
\begin{equation}
b 
= |e a^2 B|
.\end{equation}
As quantized magnetic fields on a lattice will satisfy Eq.~\eqref{eq:quantization}, 
we have
$b = 6 \pi |n_\Phi| / N_L^2$, 
where 
$N_L$ 
is the number of spatial sites. 
Typically one needs at least four values of the magnetic field strength to extract a polarizability with confidence, 
so 
$(n_\Phi)_{\text{max}} = 4$. 
The number of spatial sites must be at least
$N_L = 32$
to have pertubatively small fields compared to hadronic scales with a typical lattice spacing of
$a = 0.1 \, \texttt{fm}$. 
Thus in practice, 
we are limited to 
$b \lesssim 1/10$, 
and an expansion in 
$b$ 
is well justified.

When we are near the continuum limit, 
the wave-function of the lowest Landau level, 
$\psi^{(0)}_0 (x_2)$ in Eq.~\eqref{eq:Gauss},
restricts the coordinate to values  
$x_2 \lesssim  | e B |^{-\frac{1}{2}}$. 
Thus when we series expand the potential
$V$
about the continuum limit, 
$V = \sum_{j=0}^\infty V_{2j}$, 
the 
$j$-th 
term 
\begin{equation}
V_{2j} 
\propto 
\frac{1}{a^2} (e a B x_2)^{2 j+2} 
\lesssim
b^j \, V_0
,\end{equation}
is suppressed by 
$j$ 
powers of 
$b$. 
The kinetic operator 
$T$
can also be expanded near the continuum limit, 
$T = \sum_{j = 0}^\infty T_{2j}$. 
In the lowest Landau level, 
the momentum 
$p_2$
is also restricted, 
$p_2 \lesssim | e B|^{\frac{1}{2}}$
on account of the canonical transformation. 
Consequently, 
the 
$j$-th 
term in the expansion
\begin{equation}
T_{2j} 
\propto
\frac{1}{a^2} (a p_2)^{2 j + 2}
\lesssim
b^j \, T_0
,\end{equation}
is also suppressed by 
$j$ 
powers of 
$b$. 
Hence we address the leading-order effects of discretization by treating 
$T_2 + V_2$
as a perturbation of the continuum result.

The explicit form of the expansion to second order is
\begin{equation}
T_2
+
V_2 
=  
- \frac{1}{12 a^2} 
\left[
(a \hat{p}_2 )^4
+
b^4 
\left( \hat{x}_2 / a \right)^4
\right] \label{eq:the}
.\end{equation}
The computation now amounts to basic Rayleigh-Schr\"odinger perturbation theory. 
We compute the leading correction to the energy eigenvalue, as well as the wave-function. 
Writing the eigenvalue of the lowest lattice Landau level as
\begin{equation}
\lambda_0
= 
\lambda^{(0)}_0
+ 
\lambda^{(1)}_0
+ 
\cdots,
\end{equation}
with 
$a^2 \lambda^{(0)}_0 = b$, 
we have at once
\begin{equation} \label{eq:lambda0one}
a^2 
\lambda_0^{(1)}
=
- \frac{b^2}{8}
.\end{equation}
Having deduced this correction, 
we can write down the perturbative expansion in 
$b$ 
of the the energy eigenvalue 
$\mathcal{E}_0$
in Eq.~\eqref{eq:energy}. 
Given in lattice units, 
we find
\begin{equation}
a^2 \mathcal{E}_0^2
= a^2 M^2 + b -  \left( \frac{1}{8} + \beta  \right) b^2 + \mathcal{O} (b^3)
,\end{equation}
where the 
$\beta$
term depends on the magnetic polarizability
$\beta_M$
in the form
\begin{equation}
\beta
=
a M \frac{\beta_M}{\alpha \, a^3}
,\end{equation}
with the fine-structure constant
$\alpha = \frac{e^2}{4 \pi}$. 
Polarizabilities of hadrons have values of natural size in units of  
$10^{-4} \, \texttt{fm}^3$, 
although one expects smaller values for polarizabilities at larger-than-physical pion masses. 
For a physical mass pion,
$M = m_\pi$, 
and a lattice spacing of 
$a = 0.1 \, \texttt{fm}$, 
we expect 
$\beta \sim 1$.
From this value, 
we see that the discretization correction to the energy of the lowest Landau level will affect the extracted value of the polarizability by
$\sim 10$--$20 \%$ 
if ignored.%
\footnote{
With respect to finite lattice spacing effects, 
we must also remark that the electromagnetic current receives 
$\cO(a^2)$ 
corrections. 
Such discretization corrections have nothing to do with Landau levels, 
and are thus present even for neutral hadron correlation functions. 
Discretization corrections to the polarizability due to the electromagnetic current are proportional to
$(a \Lambda_{\text{QCD}})^2$;
which, 
on a lattice with 
$a = 0.1 \, \texttt{fm}$, 
is expected to be
$\lesssim 4 \%$
for 
$\Lambda_{\text{QCD}} \lesssim 400 \, \texttt{MeV}$. 
}

\begin{figure}
\epsfig{file=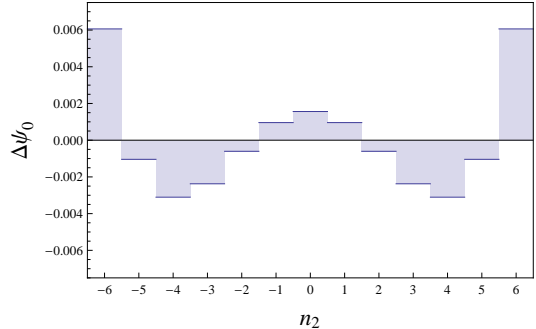,width=0.425\textwidth}
\caption{
Discretization corrections to the 
coordinate wave-function of the lowest Landau level. 
Shown as a function of 
$n_2$
is the relative difference
$\D \psi_0 ( a n_2)$
defined in 
Eq.~\eqref{eq:reldiff}.
In lattice units, 
the magnetic field strength is taken to be
$b = 0.1$. 
Values this size and smaller are required in practice.
}
\label{f:WVFN}
\end{figure}

To investigate the discretization corrections to the ground-state wave-function, 
we similarly write the coordinate wave-function as a perturbative expansion about the continuum limit
\begin{equation}
\psi_0 (x_2)
= 
\psi_0^{(0)} (x_2)
+ 
\psi_0^{(1)} (x_2)
+ 
\cdots
,\end{equation}
and we find
\begin{equation} \label{eq:wvfn}
\psi_0^{(1)} (x_2)
=
\frac{b}{16 \sqrt{6}} 
\psi_4^{(0)} (x_2)
.\end{equation}
For magnetic field strengths of size
$b \lesssim 0.1$, 
the correction to the wave-function
$\psi_0^{(1)} (x_2)$ 
in 
Eq.~\eqref{eq:wvfn} 
is quite small. 
We can exhibit this by considering the difference in the ground-state wave-function relative to the continuum one,
\begin{equation} \label{eq:reldiff}
\D
\psi_0
(x_2)
= 
\frac{\psi_0(x_2) - \psi_0^{(0)}(x_2)}{\psi_0^{(0)} (x_2)}
.\end{equation}
This relative difference is plotted in 
Fig.~\ref{f:WVFN} 
for the value
$b = 0.1$.
In practice, 
magnetic fields of this size and smaller are required to probe the magnetic polarizability of the 
$\phi$. 
From the figure, 
we see that corrections to the continuum wave-function will be less than
$1 \%$
for such magnetic fields.

%
%
%
%

Notice that the relative correction actually grows as a function of 
$n_2$
because 
it is proportional to a Hermite polynomial, 
namely
$H_4 ( \sqrt{b} \, n_2)$.
While this polynomial is unbounded, 
large corrections only occur when the wave-function is exponentially small. 
For example, 
to get a 
$10\%$ 
discretization correction to the wave-function in a field of 
$b = 0.1$, 
one must go out to a distance 
$n_2 \sim 9$
from the origin. 
But the ratio of the wave-function at this point to its value at the origin is
$\psi_0^{(0)} (a n_2) / \psi_0^{(0)} (0) \sim 10^{-2}$.
Because a determination of the correlation function itself to much better than 
$1\%$ 
is not likely, 
the resulting discretization correction to the wave-function can be safely neglected. 
Said another way,  
plots of 
$\psi_0^{(0)}(x_2)$
and 
$\psi_0^{(0)}(x_2) + \psi_0^{(1)}(x_2)$
for 
$b = 0.1$
are indistinguishable, 
unless we zoom in on the tails of the Gau{\ss}ian.

In practice, 
one will need to verify the lattice form of the dispersion relation for the particle of interest in order to address the effects of discretization. 
While the form of the dispersion relation need not be exactly that assumed in Eq.~\eqref{eq:disp}, 
our power-counting arguments readily generalize to other forms. 
For small values of 
$b$, 
only the 
$\cO(a^2)$
corrections to the proper-time Hamiltonian will be required
\begin{equation} \label{eq:blah2}
\D H = - \frac{1}{12 a^2} \left[ C_1  ( a \hat{p}_2 )^4 + C_2 \, b^4 (\hat{x}_2 / a)^4 \right]
.\end{equation}
This differs from Eq.~\eqref{eq:the} by the introduction of two arbitrary coefficients, 
$C_1$ and $C_2$. 
The coefficient 
$C_1$
can be determined numerically by studying the deviation of the dispersion relation from the continuum one, 
\begin{equation} \label{eq:expand}
E^2 = M^2 + \vec{p} \, {}^2 - \frac{a^2 C_1}{12}  \sum_{j=1}^3 (p_j)^4 + \cdots
,\end{equation}
where we have made use of cubic invariance, 
and eliminated the term proportional to 
$(\vec{p} \, {}^2)^2$
by making a field redefinition. 
The coefficient 
$C_2$ 
appearing in Eq.~\eqref{eq:blah2} 
is not an independent parameter; 
it is fixed by gauge invariance.
The effective action describing a particle with Eq.~\eqref{eq:expand} as its dispersion relation is
gauged by the replacement
$\hat{p}_j \to \hat{p}_j + e \hat{A}_j$. 
After projecting the good components of momentum,
$p_1$ 
and 
$p_3$, 
to zero in such an effective action, 
we arrive at Eq.~\eqref{eq:blah2}
with 
$C_2 = C_1$. 
Notice that additional terms, 
such as 
$(\hat{p}_2)^2 (\hat{x}_2)^2$,
are forbidden by the combination of cubic symmetry and gauge invariance.  
Thus to address the effects of discretization on the lowest Landau level, 
we perturb about the continuum with the interaction
\begin{equation}
\D H = - \frac{C_1}{12 a^2} \left[  ( a \hat{p}_2 )^4 +  b^4 (\hat{x}_2 / a)^4 \right]
.\end{equation}
The results presented above employ the value 
$C_1 = 1$, 
however, 
this coefficient should be determined non-perturbatively for the particular lattice QCD action employed in actual  calculations.  
For the pseudoscalar Goldstone mesons, 
one generally expects the breaking of rotational invariance in Eq.~\eqref{eq:expand} to be quite small~\cite{Bar:2003mh}, 
while for baryons this is not necessarily the case~\cite{Tiburzi:2005vy}.

\section{Finite Volume Effects} \label{s:FV}  %

Having established that discretization corrections can be treated in a perturbative expansion about the continuum, 
we now turn to finite volume effects. 
To study these effects, 
we consider the theory defined on a continuous torus of length 
$L$
in each of the three spatial directions, 
with the temporal extent kept infinite. 
In our analytic approach, 
we do not consider a discrete torus, 
i.e.~a finite lattice,
because the combination of both discretization and volume corrections is a doubly small effect. 
This will be verified by our numerical investigation in Sec.~\ref{s:test}. 
To determine the finite volume modifications to the projection technique, 
we must understand the behavior of charged particle two-point functions under spatial translations by 
$L$.

On a torus, 
there is a discrete magnetic translation group, 
see e.g.~\cite{AlHashimi:2008hr}, 
a few aspects of which are required to address volume corrections. 
The gauge potential 
$A_\mu$
given in Eq.~\eqref{eq:A}
is periodic up to a gauge transformation 
\begin{equation}
A_1 (x + \hat{x}_2 L)
= 
A_1(x)
+ 
\partial_1 \Lambda(x)
,\end{equation}
with all other directions strictly periodic. 
The gauge transformation function appearing above is
$\Lambda(x) = - B L x_1$. 
The corresponding gauge transformed matter field must then satisfy the magnetic periodic boundary condition
\begin{equation} \label{eq:mbpc}
\phi(x + \hat{x}_2 L)
= 
e^{ i e B L x_1} 
\phi(x)
,\end{equation} 
with the other directions periodic. 
The boundary conditions in the 
$x_1$ 
and 
$x_2$ 
directions can only be consistent if the magnetic flux through the 
$x_1, x_2$--plane
is quantized. 
We will label coordinates in this plane simply by 
$\vec{x}_\perp$. 
For down quark fields of fractional electron charge, 
the quantization condition is given in  
Eq.~\eqref{eq:quantization};
and, 
in turn,
this gives a magnetic flux quantum of 
$N_\Phi = 3 n_\Phi$
for a scalar hadron 
$\phi$
having unit charge.

To generalize the projection technique to the torus, 
we must isolate the lowest Landau level. 
The ground-state wave-function on a torus must be a sum of images of the infinite volume wave-function%
~\cite{AlHashimi:2008hr}
\begin{equation} \label{eq:psiFV}
\psi_0^{FV}(\vec{x}_\perp)
= 
\sum_\nu
\psi_0^{(0)} (x_2 + \nu L) 
e^{- 2 \pi i N_\Phi \nu x_1 / L}
,\end{equation}
in order to maintain magnetic periodicity. 
Notice that this particular finite volume wave-function corresponds to a state with zero momentum in the 
$x_1$--direction.  
With a lattice determination of the two-point function%
\footnote{
We consider the lattice to extend from 
$- \frac{L}{2}$ 
to 
$\frac{L}{2}$
in each of the spatial directions. 
This is an inessential choice, 
and serves to simplify the discussion of finite volume corrections in the winding number expansion. 
For a lattice that extends from 
$0$ 
to 
$L$
in each of the spatial directions, 
the relative change in the ground-state wave-function is exactly the same as in Eq.~\eqref{eq:FVeffect}. 
As 
$x_2$ 
now extends to 
$L$
in this case, 
there is a very large finite volume correction from the 
$\nu = -1$
image.
This is essentially the peak of the ground-state wave-function located at the origin, 
because, 
due to magnetic periodicity, 
the origin,
$x_2 = 0$, 
is equivalent to 
$x_2 = L$
up to an
$x_1$-dependent
phase. 
As such a setup is convenient for lattice calculations, 
we will adopt the asymmetric formulation in our numerical investigation below. 
}
\begin{equation} \label{eq:coord}
G^{FV}_B (\vec{x}_\perp, \tau)
= 
\int_{-\frac{L}{2}}^{\frac{L}{2}} dx_3 \,
\langle 0 | 
\Phi (\vec{x}_\perp, x_3, \tau) \Phi^\dagger (\vec{0}_\perp, 0, 0) 
| 0 \rangle_B^{FV}
\end{equation}
we then form the finite volume generalization of Eq.~\eqref{eq:Btau}
\begin{equation}
\mathcal{G}_B^{FV}
(\tau)
=
\int_{-\frac{L}{2}}^{\frac{L}{2}} d\vec{x}_\perp \,
\psi_0^{* FV}(\vec{x}_\perp) \,
G^{FV}_B (\vec{x}_\perp, \tau)
.\end{equation}
Appealing to Schwinger's proper-time trick, 
and a resolution of the identity in the infinte
$\vec{x}_\perp$-plane
in terms of cells of area 
$L^2$ 
indexed by integers
$(\nu_1, \nu_2)$, 
we find the 
$\phi$ 
contribution to the correlator indeed has a simple form
\begin{equation}
\mathcal{G}_B^{FV} (\tau) 
= 
Z_\phi \,
e^{ - E_0 \tau}
,\end{equation}
with the energy,
$E_0$,
the same as in infinite volume, 
Eq.~\eqref{eq:E0}.
\footnote{
It should be noted that there are also dynamical finite volume corrections which affect the extraction of 
$\cM^2$
that we are neglecting.
These corrections are suppressed by 
$\exp ( - m_\pi L )$;
but, 
their treatment is subtle due to the holonomy of the external field%
~\cite{Engelhardt:2007ub,Tiburzi:2008pa}. 
Dirichlet boundary conditions can be employed to remove such effects~\cite{Alexandru:2010dx}, 
however,
the cost is the introduction of different finite-size effects.
}

Taking a pion interpolating field for 
$\Phi$, 
the coordinate-space correlator in 
Eq.~\eqref{eq:coord}  
remains positive definite in an external magnetic field. 
In projecting out the ground state with the wave-function in 
Eq.~\eqref{eq:psiFV},
we thus restrict our attention to the real part of
$\mathcal{G}^{FV}_B (\tau)$, 
which depends on the real part of 
$\psi_0^{FV}(\vec{x}_\perp)$. 
Consider the change in the finite volume wave-function relative to the infinite volume one, 
\begin{equation}
\Delta \psi_0^{FV} (\vec{x}_\perp)
= 
\frac{ 
\mathfrak{Re} \left[ \psi_0^{FV} (\vec{x}_\perp) \right] 
- 
\psi_0^{(0)} (x_2)}
{\psi_0^{(0)} (x_2)}
.\end{equation}
This relative change takes the form 
\begin{eqnarray} \label{eq:FVeffect}
\Delta \psi_0^{FV} (\vec{x}_\perp)
&=& 
2 
\sum_{\nu = 1}^\infty
e^{ - \nu^2 \pi  N_\Phi }
\cos ( 2 \pi N_\Phi \nu x_1 / L)
\cosh ( 2 \pi N_\Phi \nu x_2 / L)
.\end{eqnarray}

%
%
\begin{figure}[t!]
\epsfig{file=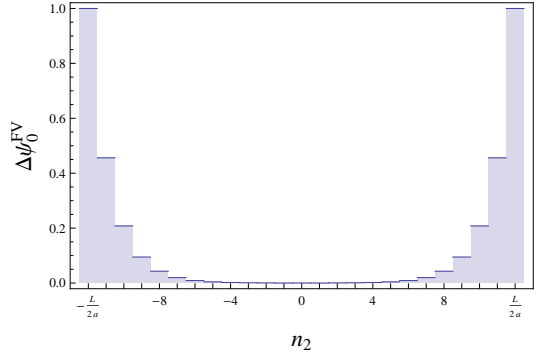,width=0.425\textwidth}
\caption{
Finite volume corrections to the coordinate wave-function of the lowest Landau level. 
The plot shows the relative difference
$\D \psi_0^{FV} (x_1 =0, x_2 = a n_2)$
defined in 
Eq.~\eqref{eq:FVeffect}
as a function of 
$n_2$.
We choose the minimal flux quantum, 
$N_\Phi = 3$,
and a relatively small lattice size,
$L = 24 a$,
for which the corresponding value of the magnetic field is
$b = 0.033$.
Finite volume corrections for 
$x_1 \neq 0$
are bounded in absolute value by those shown in the plot. 
}
\label{f:FV}
\end{figure}
%
%

Physically we expect volume corrections to be largest in the smallest magnetic field. 
The smallest magnetic field allowed on the torus has the flux quantum 
$N_\Phi = 3$. 
A plot of the relative change in the wave-function 
$\D \psi^{FV}_0 (\vec{x}_\perp)$
is shown in 
Fig.~\ref{f:FV}
for a magnetic field with flux quantum
$N_\Phi = 3$
on a lattice of size
$L = 24 a$. 
The relative change shown in the figure overwhelmingly arises from the winding number 
$\nu = 1$. 
Contributions from 
$\nu = 2$, 
for example, 
are suppressed by at least  
$e^{ - 6 \pi}$,
as we show below.
Not surprisingly, 
the largest finite volume corrections arise at the 
$x_2$-boundary 
of the lattice.

Let us carefully analyze the relative change 
$\D \psi^{FV}_0 (\vec{x}_\perp)$
by considering the contribution from a generic winding number 
$\nu$
in the series.  
The 
$x_1$-dependence 
of this contribution enters through the cosine, 
$\cos \left( 2 \pi N_\Phi \nu x_1 / L \right)$, 
and so is always bounded by unity. 
The 
$\nu^{\text{th}}$ 
term in the series depends exponentially on the magnitude of 
$x_2$
through the hyperbolic-cosine.
Because the 
$x_2$-coordinate 
is bounded, 
$| x_2| \leq \frac{L}{2}$, 
however, 
we have 
$\cosh \left( 2 \pi N_\Phi \nu x_2 / L \right) \leq \frac{1}{2} e^{ \nu \pi N_\Phi }$
to exponential accuracy.  
Thus the contribution to the finite volume effect from a generic winding number 
$\nu$
is bounded by
$e^{- \nu ( \nu - 1) \pi N_\Phi}$, 
i.e.~finite volume effects are suppressed by an exponential factor involving the magnetic flux quantum. 
This suppression occurs provided the winding number is greater than one. 
When 
$\nu = 1$, 
there is no exponential suppression in the relative change when 
$x_2$ 
approaches the boundary of the lattice.

At the 
$x_2$-boundary 
of the lattice, 
we have
\begin{equation}
\D \psi_0^{FV} \left(x_1 = 0, x_2 = \pm L/2 \right) = 1
,\end{equation}
to exponential accuracy, 
which is independent of the flux quantum 
$N_\Phi$.
While this corresponds to the maximal value for the relative change, 
we should also consider the size of the ground-state wave-function at the 
$x_2$-boundary,
because this is where the wave-function is the smallest. 
The ratio of the infinite volume wave-function at the boundary to its value at the origin is given by
\begin{equation}
\psi_0^{(0)} \left(x_2 = \pm L/2 \right) / \psi_0^{(0)} (0) = e^{ - \frac{\pi}{4} N_\Phi}
,\end{equation} 
which is remarkably independent of the lattice size 
$L$.
Unlike discretization corrections to the wave-function, 
finite volume corrections are generally non-negligible. 
For example, 
in the smallest magnetic field, 
$N_\Phi = 3$,
the boundary-to-origin ratio 
$\sim 0.1$
is not considerably less than unity
and 
one must consider the 
$\sim 100 \%$
correction to the infinite volume wave-function from the 
$\nu = 1$
image.

The independence of this result on the lattice size
$L$
brings us to our final point. 
There is a subtlety in taking the infinite volume limit. 
The finite volume correction in Eq.~\eqref{eq:FVeffect}
does not vanish in the limit 
$L \to \infty$. 
This is due to the magnetic field quantization condition. 
As 
$L$
is made larger, 
the magnetic fields allowed on the torus becomes smaller.
When the magnetic field becomes smaller, 
however,  
the ground-state wave-function spreads out in space. 
In this way, 
increasing the volume need not result in smaller finite volume effects. 
The tacit assumption in this approach to infinite volume is that the flux quantum 
$N_\Phi$
remains fixed as 
$L \to \infty$. 
This is likely to be the case in practice, 
as one will generate larger lattices and then focus on the smallest allowed values of the magnetic field on these lattices. 
To recover the infinite volume limit,
however, 
the magnetic field should be fixed
\footnote{
Actually one need not be this restrictive to recover the infinite volume limit.
As long as the flux quantum scales with some positive power 
$p$
of the lattice size, 
the infinite volume limit will be recovered.
The finite volume correction to the wave-function remains  
$\sim 100 \%$
at the 
$x_2$-boundary, 
but the size of the wave-function at the 
$x_2$-boundary 
is
$\propto \exp \left( - \frac{\pi}{4} N_\Phi \right)$
which will be exponentially small in the volume for 
$N_\Phi \propto L^p$.
}
as 
$L \to \infty$, 
which in turn requires 
$N_\Phi \propto L^2$.  
While there are still
$\sim 100 \%$ 
finite volume corrections to the wave-function from Eq.~\eqref{eq:FVeffect}, 
these occur at the 
$x_2$-boundary which is now precisely where the wave-function is exponentially suppressed in the volume, 
$\propto e^{ - \frac{1}{8} | e B| L^2}$.
With such corrections buried in the tails of the Gau{\ss}ian wave-function, 
the infinite volume limit is recovered.

\section{Numerical Test}
\label{s:test}

To demonstrate the proposed method, 
we provide a numerical test using the lattice action for a scalar particle coupled to an external magnetic field. 
Unlike our approximations above, 
the theory is rendered on a finite, 
four-dimensional Euclidean lattice with sites
$x_\mu = a n_\mu$, 
using the notation
$n_\mu = (n_1, n_2, n_3, n_4)$
to label a four-vector of integers.  
In this way, 
effects of both the discretization and finite volume are included simultaneously, 
and the analytic approach we detail above can be tested in a controlled setting.  
The scalar particle action has the form
\begin{eqnarray}
S
&=&
a^4
\sum_{\vec{n}, \vec{n}' = 0}^{N_L - 1} \,
\sum_{n_4, n'_4 = 0}^{N_T - 1}
\phi^\dagger_{n} 
\,
\cD_{n,n'}
\,
\phi_{n'} 
,\end{eqnarray}
with the length of the time direction given by 
$T = a N_T$, 
and the 
length of each of the three spatial directions given by 
$L = a N_L$. 
In our numerical investigation, 
we shall use a lattice having
$N_L^3 \times N_T = 32^3 \times 64$
sites.
We keep explicit powers of the lattice spacing only to employ consistent notation.
In the actual computation of correlation functions, 
lattice units are naturally employed. 
The action is specified by the matrix   
$\cD_{n,n'}$, 
which has the form
\begin{eqnarray}
\cD_{n,n'}
&=&
-
\frac{1}{a^2}
\sum_{\mu=1}^4
\left[
\d_{n + \hat{\mu},n'}
\, U_{\mu,n}
+ 
\d_{n, n' + \hat{\mu}}
\, U_{\mu,n'}^\dagger
- 
\frac{1}{4}
\left(
8
+ 
a^2 \cM^2
\right)
\d_{n,n'}
\right]
.
\label{eq:blah3}
\end{eqnarray}
The gauge links, 
$U_{\mu, n}$, 
are lattice site-dependent phases specified by
\begin{equation}
U_{\mu,n}
=
\exp \left( - i b  \,n_2 \, \delta_{\mu, 1} \right)
\exp \left(+ i b \, n_1 \, N_L \delta_{\mu, 2} \delta_{n_2, N_L - 1} \right)
,\end{equation}
with 
$b$
subject to the quantization condition in Eq.~\eqref{eq:quantization}. 
Written in lattice units,
we have 
$b = \frac{6 \pi }{N_L^2} n_\Phi$.
The additional factor of 
$3$
appearing in the quantization condition reflects that, 
even though we choose the scalar to have unit charge, 
we imagine it composed from quarks having fractional electric charges. 
In the continuum and infinite volume limits, 
the links generate a uniform magnetic field. 
The parameter 
$\cM$
appearing in the action
is that in 
Eq.~\eqref{eq:M},%
\footnote{
Alternatively 
we could include the polarizability using the discretized form of the $s$-wave and $d$-wave couplings in 
Eq.~\eqref{eq:effact}.
This approach leads to additional terms of order 
$b^4$
that are negligible in our study. 
}
however, 
we fix the value of the polarizability to the expected size, 
namely
\begin{equation}
a^2 \cM^2 
= 
a^2 M^2
-  b^2
,\end{equation}
namely 
$\beta \equiv 1$. 
The scalar particle is thus point-like with a non-minimal coupling to the external field which is its magnetic polarizability.

Using the action specified by Eq.~\eqref{eq:blah3}, 
we can numerically compute various two-point correlation functions of the scalar particle. 
The two-point functions that we calculate
can be written in a generic form
\begin{equation}
G_B^{(\alpha)}
(\tau)
=
\sum_{\vec{n}=0}^{N_L - 1}
g^{(\alpha)}_B (n_1, n_2)
\langle 0 | \phi^{\phantom{\dagger}}_{\vec{n}, \frac{\tau}{a}} \, \phi^\dagger_{\vec{0}, 0} | 0 \rangle
,\end{equation}
with 
$g^{(\alpha)}_B$
as functions that  possibly depend on the lattice sites. 
We consider three such functions. 
The first corresponds to the zero momentum projection, 
\begin{equation}
g^{(1)}_B = 1
,\end{equation} 
as in Eq.~\eqref{eq:dumb}. 
The second corresponds to a na\"ive projection of the lowest Landau level, 
and is specified by 
\begin{equation}
g^{(2)}_B (n_2) 
= 
\psi_0^{(0)} ( a n_2)
+ 
\psi_0^{(0)} ( a n_2 - L )
.\end{equation} 
Here we employ the continuum wave-function for the lowest Landau level, 
as in Eq.~\eqref{eq:Btaulatt}. 
Because the scalar source is not located symmetrically in the middle of the lattice, 
we include the na\"ive image. 
Given the properties of the magnetic translation group, 
however,
the correct image from Eq.~\eqref{eq:psiFV} requires the phase factor, 
$\exp \left(- i b N_L n_1 \right)$, 
which reflects the holonomy of the gauge field. 
By omitting this phase factor, 
the function 
$g_B^{(2)}$ 
only depends on 
$n_2$, 
rather than on both 
$n_1$ 
and 
$n_2$. 
Finally we choose a third function, 
which corresponds to our best guess based on our analytic study.
This function depends on both 
$n_1$
and
$n_2$
in the form
\begin{equation}
g_B^{(3)} (n_1, n_2)
=
\psi_0^{(0)} ( a n_2)
+ 
e^{-i b N_L n_1}
\psi_0^{(0)} ( a n_2 - L )
+
e^{i b N_L n_1} 
\psi_0^{(0)} (a n_2 + L)
+
e^{-2 i b N_L n_1}
\psi_0^{(0)} (a n_2 - 2 L)
.\end{equation}
Weighting the two-point correlator with this function corresponds to projection of the lowest Landau level using the continuum wave-function,
and the first non-trivial images including the proper phases. 
We argued above that additional magnetic-periodic images are exponentially suppressed, 
and that the leading discretization correction to the wave-function should similarly be small. 
For these reasons, 
we deem 
$g_B^{(3)}$
to be our best guess. 
We could obviously improve the projection should the data require it.

%
%
\begin{figure}
\epsfig{file=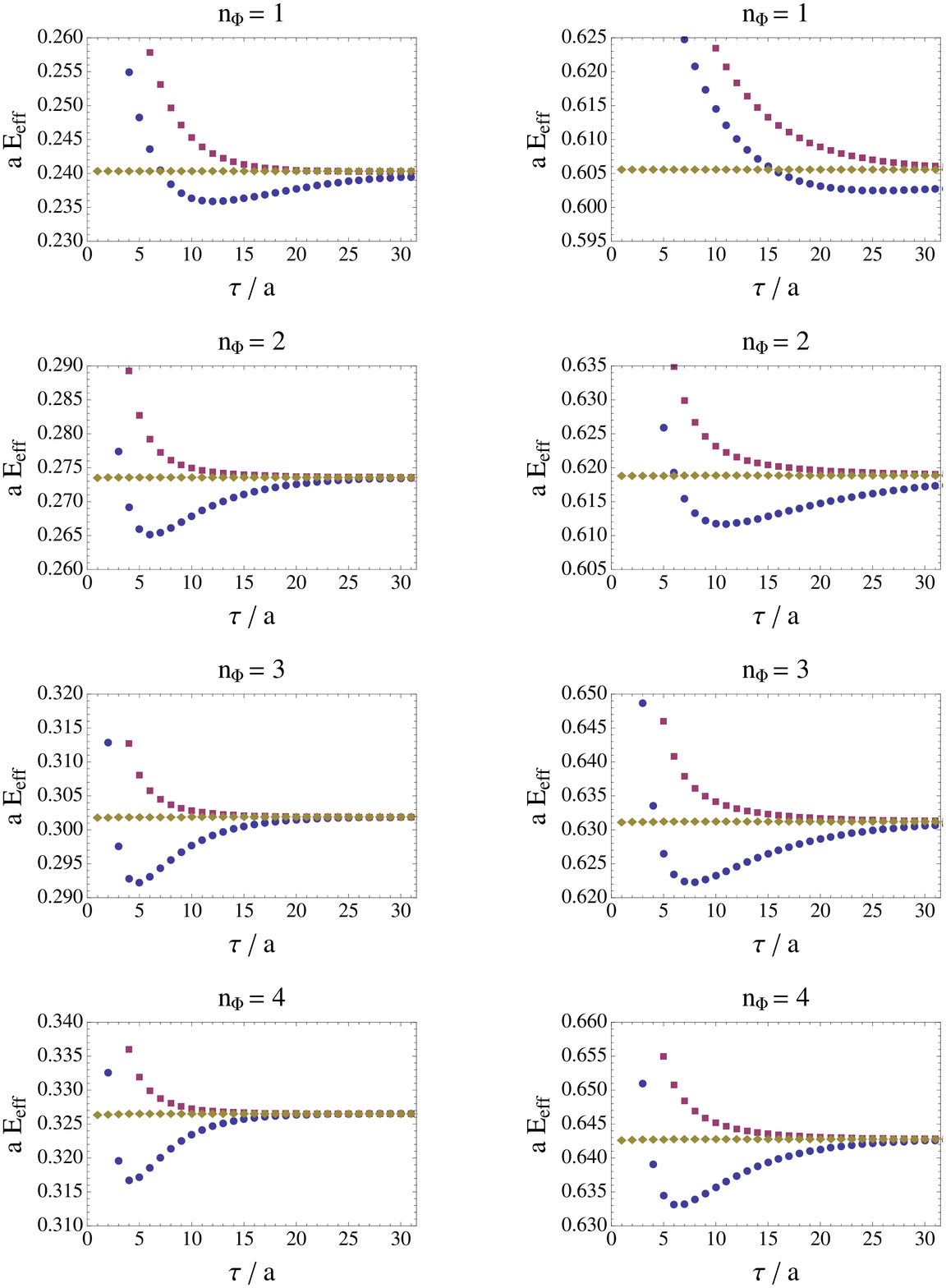,width=\textwidth}
\caption{Effective energy plots for a light scalar particle (left panels) and a heavy scalar particle (right panels). 
Shown are the effective energies as a function of time 
$\tau$
for the lowest four accessible magnetic flux quanta, 
$n_\Phi =1$--$4$. 
The squares, circles, and diamonds show the effective energy determined from the correlators
$G_B^{(1)}$,
$G_B^{(2)}$,  
and
$G_B^{(3)}$, 
respectively. 
Due to temporal periodicity, 
we only show results up to time
$\tau = (T- a) / 2$.
}
\label{f:test}
\end{figure}
%
%

We numerically computed the three correlation functions, 
$G^{(\alpha)}_B (\tau)$
for 
$\alpha = 1$--$3$, 
for the lowest four magnetic flux quanta, 
$n_\Phi = 1$--$4$. 
From the correlators, 
we determine the effective energy from the ratio of their values on neighboring time slices. 
Specifically we take into account periodicity in time by solving the equation
\begin{equation}
\frac{\mathfrak{Re} 
\left[ 
G_B^{(\alpha)}(\tau + a)
\right]}
{\mathfrak{Re} 
\left[
G_B^{(\alpha)}(\tau)
\right]
} 
= 
\frac{
\cosh \left[ E^{(\alpha)}_{\text{eff}} \left( \frac{T}{2} - (\tau + a) \right) \right]
}
{
\cosh \left[ E^{(\alpha)}_\text{eff} \left( \frac{T}{2} - \tau\right) \right]
}
,\end{equation} 
for 
$E^{(\alpha)}_{\text{eff}}$
as a function of 
$\tau$. 
In 
Fig.~\ref{f:test}, 
we plot the resulting effective energies for two choices of the mass parameter. 
We consider a light scalar of mass
$a M = 0.2$, 
and a heavy scalar of mass
$a M = 0.6$. 
If we imagine the lattice spacing to be 
$a = 0.1 \, \texttt{fm}$, 
then the light scalar corresponds to a particle with mass 
$\sim 400 \, \texttt{MeV}$, 
while the heavy scalar corresponds to a particle with mass
$\sim 1.2 \, \texttt{GeV}$. 
Thus the case of the light scalar we imagine to be relevant for a heavier-than-physical charged pion, 
while that of the heavy scalar is more relevant for the proton, 
minus the spin degrees of freedom of course.

The figure exhibits trends discussed above. 
Despite describing only a single hadron, 
the zero-momentum correlator, 
$G_B^{(1)}(\tau)$, 
requires long times to exhibit a plateau in the effective energy. 
In a magnetic field, 
the zero-momentum state of the
$\phi$
is a superposition of Landau levels, 
each having a different energy. 
The figure shows that a plateau is reached faster in Euclidean time for the light scalar particle compared to the heavy scalar. 
Additionally for a fixed mass, 
a plateau is reached earlier in stronger magnetic fields. 
Both of these features are easily explained in terms of the spacing between Landau levels. 
The splitting between adjacent Landau levels is given by
$a \, \Delta E = b / (a M)$. 
Thus the larger the mass
(or the smaller the magnetic field), 
the more challenging the spectroscopy. 
In a lattice QCD calculation, 
excited-state hadrons will also contribute to the two-point function. 
For the nucleon, 
the signal-to-noise problem will restrict one to fitting correlator data at early times. 
From the plots for the heavy scalar particle, 
restricting the zero-momentum correlator to times 
$\tau / a < 16$
will not allow one to determine the energy to better than several percent in the smallest two magnetic fields. 
As the effect from the magnetic polarizability is itself at most a few percent effect, 
the zero-momentum projection will not be practicable to extract the polarizability.  
For the pion, 
one is not restricted by statistical noise, 
and data at larger time separations can be utilized. 
From the plots for the light scalar particle, 
extracting the energy at the percent level or better from the long-time behavior of the correlator should not pose any problem. 
We should note, 
however, 
that the energy shift due to the magnetic field is already at the 
$20 \%$
level for the smallest magnetic field. 
Even though the Landau levels do not look problematic for the light scalar, 
one may not be in the regime of sufficiently small magnetic fields.  
On a larger lattice,  
of size 
$48^3 \times 96$, 
for example,
one will have access to smaller magnetic fields, 
but then the Landau level spacing for the light scalar will be comparable to that of the heavy scalar on 
$32^3 \times 64$ 
lattices.

The Landau levels of charged particles present a systematic effect that can be completely removed by the projection method. 
To this end, 
we consider the na\"ive projection used to form the 
$G_B^{(2)}(\tau)$
correlation function. 
Fig.~\ref{f:test} 
shows that the na\"ive projection does not offer much improvement over using the zero-momentum correlator. 
While na\"ive projection would work in the infinite volume and continuum limits, 
the effective energy exhibits non-trivial 
$\tau$-dependence, 
which is indicative of contributions from multiple energy eigenstates rather than the lowest lattice Landau level. 
The effective energies do not decrease monotonically, 
moreover, 
which suggests a mismatch of phase factors in the spectral decomposition of the two-point function. 
Fortunately this situation can be remedied.  
Using the analytic observations from above,
we know how to improve the projection technique, 
and accordingly form the best-guess correlator, 
$G_B^{(3)}(\tau)$. 
From the plots of the effective energy resulting from the best-guest correlator, 
we see that it is precisely what is needed to remove the systematic effect due to Landau levels. 
The effective energies exhibit a plateau immediately in Euclidean time. 
In fact, 
the time variation of the effective energies shown is at most a few parts in $10^{-4}$, 
and is an order of magnitude greater than the precision of lattice data required to extract the polarizability, 
see Table~\ref{t:compare}. 
The value of the effective energy extracted from the best-guess correlator lines up with the analytic expectation, 
\begin{equation} \label{eq:analytic}
4 \sinh^2 
\left(
a E / 2 
\right) 
= 
(aM)^2 
+ 
b 
- 
\left( \frac{1}{8} + 1
\right) 
b^2
,\end{equation}
where we have taken into account the temporal discretization. 
This analytic value, 
moreover, 
matches with the numerical determination of the lowest eigenvalue of the action, 
which should be considered the exact solution to the lattice Landau level problem. 
With the lowest lattice Landau level confidently isolated, 
one can fit lattice data at earlier times with the only excited-state contamination arising from exited-state hadrons, 
just as in the absence of magnetic fields.

\begin{table}[t]
\caption{\label{t:compare} 
Effective energies extracted from the best-guess correlator compared to the exact results. 
For the best-guess effective energy, 
we choose the 
\emph{worst} 
possible value, 
namely that calculated from the first two time slices. 
The exact value quoted is the numerically determined smallest eigenvalue; 
and, 
in each case,  
agrees to five significant digits with the analytic value obtained from Eq.~\eqref{eq:analytic}. 
The relative difference
$\D$ 
is defined by
$\D = |E_\text{eff}^{(3)} - E_\text{exact} | / E_\text{exact}$. 
The zero-field values are included for reference. 
Tabulated values for 
$\frac{1}{2} b (a M)^{-2}$
give the relative shift in energy due to the zero-point Landau energy, 
while values of
$\frac{1}{2} b^2 (a M)^{-2}$
give the relative shift in energy due to the value chosen for the polarizability, 
\emph{cf}.~Eq.~\eqref{eq:analytic}. 
}
\begin{center}
\begin{tabular}{cccc|cc|c}
$\quad a M \quad$ 
&
$\quad n_\Phi \quad$
&
$\quad \frac{1}{2} b (a M)^{-2} \quad$
&
$\frac{1}{2} b^2 (a M)^{-2} \, [10^{-4}] $
&
$\qquad a E_{\text{eff}}^{(3)} \qquad$
& 
$\qquad E_\text{exact} \qquad$
&
$\quad \D \,  [ 10^{-4}] \quad$
\\
\hline
\hline
$0.2$
&
$0$
&
$0$
&
$0$
&
$0.19967$
&
$0.19967$
&
$< 0.1$
\\
$0.2$
&
$1$
&
$0.23$
&
$42$
&
$0.24028$
&
$0.24031$
&
$1.2$
\\
$0.2$
&
$2$
&
$0.46$
&
$170$
&
$0.27347$
&
$0.27354$
&
$2.6$
\\
$0.2$
&
$3$
&
$0.69$
&
$380$
&
$0.30171$
&
$0.30183$
&
$3.9$
\\
$0.2$
&
$4$
&
$0.92$
&
$680$
&
$0.32630$
&
$0.32647$
&
$5.1$
\\
\hline
$0.6$
&
$0$
&
$0$
&
$0$
&
$0.59135$
&
$0.59135$
&
$<0.1$
\\
$0.6$
&
$1$
&
$0.026$
&
$4.7$
&
$0.60552$
&
$0.60554$
&
$0.4$
\\
$0.6$
&
$2$
&
$0.051$
&
$19$
&
$0.61873$
&
$0.61880$
&
$1.2$
\\
$0.6$
&
$3$
&
$0.077$
&
$42$
&
$0.63105$
&
$0.63118$
&
$2.1$
\\
$0.6$
&
$4$
&
$0.10$
&
$75$
&
$0.64254$
&
$0.64274$
&
$3.1$
\\
\hline \hline
\end{tabular}
\end{center}
\end{table}

\section{Summary} \label{s:summy}%

Above we explore the correlation functions of charged spinless hadrons in external magnetic fields. 
For magnetic fields that are small compared to the hadron's mass, 
$| e B| / M^2 \ll 1$, 
the hadron's magnetic properties can, 
in principle,  
be measured from lattice QCD simulations. 
Such computations, 
in practice,  
require modification of standard lattice spectroscopic techniques. 
This modification is necessitated by closely spaced Landau levels that cannot be cleanly resolved from the long-time 
limit of Euclidean correlation functions. 
To handle this complication, 
we develop a projection technique to isolate the lowest Landau level. 
The technique requires a modified two-point correlation function, 
Eq.~\eqref{eq:Btau}, 
that depends on the coordinate-space 
wave-function of the lowest Landau level. 
To put the technique into practice, 
one needs to write out the coordinate-time dependence of the correlator data. 
This procedure allows one to later convolve the lattice data with the ground-state wave-function, 
as one may wish to use a continuum wave-function, or include effects from discretization and the finite volume.

We investigate the effects of discretization on the lowest Landau level by assuming a form for the discrete action of charged scalar field. 
For typical lattice sizes, 
the effects of discretization can be treated in perturbation theory about the continuum limit. 
The perturbative correction to the eigenvalue and wave-function of the lowest Landau level are given in 
Eqs.~\eqref{eq:lambda0one} 
and \eqref{eq:wvfn}, 
respectively. 
The discretization correction to the energy could affect the extraction of the polarizability at the 
$\sim 10$--$20\%$ 
level. 
On the other hand, 
the discretization corrections to the wave-function of the lowest Landau level are shown to be negligible. 
Effects of boundary conditions are also considered. 
The wave-function for the lowest Landau level on a torus, 
Eq.~\eqref{eq:psiFV}, 
is used to account for finite volume effects. 
While the size of finite volume corrections is generally set by an exponentially small factor involving the magnetic flux quantum
and the winding number 
$\nu$, 
there is no exponential suppression from contributions with 
$\nu = 1$. 
We show that such finite volume corrections 
near the lattice boundary can be important even as the lattice volume increases. 
This pernicious effect owes to the magnetic flux quantization required on a torus. 
Caution must be exercised in assessing the size of finite volume corrections to the wave-function even on large lattices.
Nonetheless, 
we show how to determine the discretization and volume corrections to the projection technique, 
and thereby how to isolate the lowest Landau level from lattice 
correlation function 
data.

To test the approach, 
we implement the method for a point-like scalar particle coupled to a magnetic field.
We compare the case of a light scalar in a magnetic field to that of a heavy scalar on a lattice of size
$32^3 \times 64$. 
As expected, 
the heavy scalar is more susceptible to a pileup of Landau levels, 
and the zero-momentum correlator cannot be used to determine the ground-state energy to better than several percent in the smallest magnetic fields. 
Because the extraction of the polarizability requires at least percent-level accuracy, 
the Landau levels must be treated directly. 
The projection technique is shown to isolate the lowest lattice Landau level efficiently when the first non-trivial magnetic periodic images are included. 
Results of our numerical investigation line up precisely with the expectations from our analytic study. 
While the correlation functions for the light scalar particle do not suffer a severe problem from the Landau levels, 
the strength of the magnetic field is not necessarily perturbatively small in this case. 
Nonetheless, 
the systematic effect from Landau levels can be eliminated with the projection technique, 
and allows one to utilize correlator data at earlier times.

Further work will allow one to extend the technique. 
Inclusion of spin degrees of freedom is necessary to handle proton correlation functions. 
As the typical magnetic field strengths are semi-relativistic, 
one must account for the fact that the current and anomalous magnetic moment operators in the effective proton Hamiltonian 
do not commute. 
A method to treat Landau levels and Zeeman splittings in a relativistic context should be sought. 
Another avenue for investigation is to consider the effects of discretization in external electric fields.
The functional form of single-particle correlation functions employed by%
~\cite{Detmold:2009dx,Detmold:2010ts}
is modified by these effects. 
Fortunately these studies employed anisotropic lattices with a temporal lattice spacing of
$a_t = 0.035 \, \texttt{fm}$. 
On isotropic lattices, 
however, 
discretization effects might be important. 
One can use the methods established here to investigate the finite lattice spacing corrections 
to the extraction of the electric properties of hadrons using background field correlators. 
Finally, 
although this work deals with weak magnetic fields, 
there has been considerable interest in studying lattice QCD in strong magnetic fields due to applications relevant for heavy-ion collisions, see, for example,~\cite{Buividovich:2008wf,Buividovich:2009wi,D'Elia:2010nq,D'Elia:2011zu,Braguta:2011hq,Yamamoto:2011gk,Bali:2011qj,Bali:2012zg}. 
As stronger magnetic fields are more susceptible to discretization effects, 
one could use the methods developed here to investigate finite lattice spacing corrections in chiral perturbation theory 
with strong-field power counting, 
$| e B| \sim m_\pi^2$, 
or in perturbative QCD.

Sustained progress in lattice QCD has lead, 
in particular,
to the generation of gauge ensembles on large physical volumes. 
Such volumes will soon permit the study of QCD properties in uniform magnetic fields that are perturbatively weak compared to hadronic scales. 
This opens up the possibility to study the magnetic properties of hadrons using the external field method.  
With the projection technique developed here, 
charged hadrons can be studied by confronting the Landau levels directly. 
We look forward to exploratory lattice studies employing the technique.


\begin{acknowledgments}
Work supported in part by a joint CCNY--RBRC fellowship, 
a PSC-CUNY award, 
and by the U.~S.~National Science Foundation, 
under Grant No.~PHY$12$-$05778$.
\end{acknowledgments}

\appendix

\bibliography{hb}

\end{document}